\documentclass[%
reprint,
twocolumn,
showpacs,preprintnumbers,
amsmath,amssymb,
aip,
nofootinbib,
longbibliography, 
floatfix,
]{revtex4-2}

\usepackage{graphicx}
\usepackage{epstopdf}
\usepackage{dcolumn}
\usepackage{bm}
\usepackage{hyperref}
\usepackage{color,soul}
\usepackage{lipsum}  
\usepackage[export]{adjustbox}
\usepackage{units}

\newcommand{\ie}{\textit{i.e.,}\ }

\newcommand{\md}{\mathrm{d}}

\emergencystretch=\maxdimen
\begin{document}


\title{Generating optical supercontinuum and frequency comb in tenuous plasmas }
\author{Kenan Qu}
\author{Nathaniel J. Fisch}
\affiliation{%
	Department of Astrophysical Sciences, Princeton University,  Princeton, New Jersey 08544, USA \looseness=-1
}%

\date{\today}

\begin{abstract}
There are several mechanisms by which the frequency spectrum of a laser broadens when it propagates at near-relativistic-intensity in tenuous
plasma. Focusing on one dimensional effects, we identify two strong optical nonlinearities, namely, four-wave mixing (FWM) and forward Raman scattering (FRS), for creating octave-wide spectra.  FWM dominates the interaction when the laser pulse is short and intense; Its combination with phase modulation produces a symmetrically broadened supercontinuum. FRS dominates when the laser pulse is long and relatively weak; It broadens the laser spectrum mainly towards lower frequencies and produces a frequency comb. The frequency chirping combined with group velocity dispersion compresses the laser pulse causing higher peak intensity. 
\end{abstract}

\maketitle

\section{Introduction}
Wide-band laser spectra~\cite{Thomson_pf1974, Mostovych_prl1987, Marozas_prl2018} with high intensities are capable of overcoming plasma instabilities, such as Raman/Brillouin scattering~\cite{McKinstrie_pfb1992, Antonsen_prl1992, Antonsen_pfb1993, Mori_prl1994, Decker_1996PoP, Decker_pop19962, lezhnin2020suppression} and filamentation~\cite{Antonsen_prl1992, Antonsen_pfb1993}, and enable efficient laser power delivery~\cite{Edwards_pop2017, Zhao_pop2017, Palastro_pop2018}. Numerical simulations shows that they can propagate over an extended distance through plasmas without incurring significant pulse distortion~\cite{Solodov_pop2003, Kalmykov_pre2008, Kalmykov_2009} or absorption~\cite{Tzeng_prl1996, Edwards_pop2017, Follett_prl2018}. Hence, they are useful in numerous laser-plasma applications, such as laser-plasma accelerator~\cite{LPA_rmp2009}, inertial confinement fusion~\cite{ICF_pop2015, Palastro_pop2018}, and EUV radiation~\cite{xray_rmp2013}. But traditional methods using nonlinear optical crystals cannot efficiently produce wide-band spectra at high power~\cite{thermal_ao1970, damage_prl1995}.

Actually, plasma  can be used to broaden the laser spectrum because of its strong optical nonlinear susceptibility and high thermal resistance. It has been theoretically proposed~\cite{Cohen_prl1972, Rosenbluth_prl1972, Salomaa_1986, KarttunenPRL1986, Gibbon_prf1990, Kalmykov_2005PRL, Kalmykov_2006PRE} that a dual-color continuous laser wave can broaden its spectrum in plasma through cascades of forward Raman scattering (FRS)~\cite{Salomaa_1986, KarttunenPRL1986}. The plasma electron waves couple laser components that are detuned by the plasma frequency. Hence, a pair of continuous lasers can be converted into a wide-band spectrum with discrete spikes, \ie an optical frequency comb~\cite{OFC_rmp2003, OFC_prl2004, OFC_nature2002, OFC_nature2019}. If one of the laser pulses is at relativistic intensity ($\gtrsim\unit[10^{18}]{W cm^{-2}}$ at $\sim \unit[1]{\mu m}$ wavelength), a frequency comb could also be generated in the plasma wakefield~\cite{Yu_NC2016}.  

In this paper, we revisit the cascaded broadening of the laser spectrum in plasmas and point out a new regime dominated by four-wave mixing (FWM)~\cite{FWM_ol1979, FWM_1991, Malkin_pre2020, Malkin_pre2020(2)}. The FWM-dominated process produces a wide-band continuous spectrum without discreteness, \ie an optical supercontinuum~\cite{Supercontinuum_rmp2006, Supercontinuum_pt2013}. We compare the FWM and FRS processes to show that the cascade generates a supercontinuum when the laser pulse is short relative to a few plasma wave periods and has near relativistic intensity, and otherwise it generates a frequency comb. Short pulse duration and high intensity causes strong phase modulation and chirping. It flattens each frequency sideband and creates a continuous spectrum. 
Given sufficient plasma length, either the supercontinuum or frequency comb can reach almost a full electromagnetic spectrum spanning multiple octaves.

To create the wide-band spectrum, a pair of co-propagating laser pulses with detuning of the plasma frequency is sent into plasma. They beat to create a plasma Langmuir wave~\cite{Cohen_prl1972, Rosenbluth_prl1972} if FRS dominates, or an electron mass perturbation (a virtual phonon) if FWM dominates. Both the plasma wave and the virtual phonon scatter the laser  and broadens the spectrum at multiples of the plasma frequency. 
From a quantum point of view, a photon can either split into a lower-frequency photon and a phonon/virtual phonon, or convert itself into a higher frequency photon by combining with a phonon/virtual phonon. Since the frequency downconversion process takes place at a higher interaction rate, the phonon/virtual phonon number grows and the spectrum expands. The process can terminate due to wavevector mismatch near plasma frequency before exiting the plasma medium.

The plasma wave excited through FRS has a phase velocity near the speed of light but no group velocity. Zero group velocity means that the plasma wave propagates backward in the laser frame, causing energy transportation from the laser front to the tail. The growing plasma wave amplitude leads to a wider spectrum in the laser pulse tail than the front. Due to the energy consumption by the plasma wave, the laser spectrum shows overall frequency downshift.

FWM is a parametric process which does not excite a plasma wave and thus conserves the total electromagnetic energy. The virtual phonon amplitude does not grow or decay and is solely determined by the instantaneous laser waves.  Hence, the virtual phonon has the maximum amplitude at the laser peak where the widest spectrum broadening takes place. With conserved laser energy and photon number, the spectrum is broadened symmetrically besides the input laser frequency. The short pulse duration and high intensity combine to produce a strong chirp to stretch each frequency sideband and form a supercontinuum.  

This paper focuses on the 1D laser pulse evolution to compare the FWM and FRS processes. It is organized as follows: In Sec.~\ref{sec:model}, we model the laser propagation problem and explain the laser nonlinearities. In Sec~\ref{sec:regimes}, we identify two different interaction regimes dominated by FRS and FWM, respectively, and find the growth of the plasma wave and virtual phonon. In Sec.~\ref{sec:scaling}, we find the scaling laws of the frequency bandwidth in each regime. In Sec.~\ref{sec:envelope}, we analyze the temporal envelope evolution of the laser pulse. In Sec.~\ref{sec:PIC}, we demonstrate using PIC simulations the generation of a supercontinuum and a frequency comb. In Sec.~\ref{sec:concl}, we present our conclusions.

\section{model} \label{sec:model}
We consider cold plasma which responses to the laser field only through the electromagnetic potentials and relativistic effects. The laser pulse evolution is described by the coupled laser plasma equations in the 1D form~\cite{Antonsen_pfb1993, LPA_rmp2009}
\begin{align}
	&(\partial_{tt} - c^2\partial_{zz})a = \frac{\omega_p^2}{\gamma} \frac{n}{\bar{n}} a 
	\cong  \omega_p^2 (1+\tilde{n} - \frac{a^2}{2} ) a  , \label{1} \\
	&(\partial_{tt} + 2\nu \omega_p + \omega_p^2) \tilde{n} = \frac{c^2}{2} \partial_{zz}a^2, \label{2}
\end{align}
where $a$ is the dimensionless vector potential normalized to the laser intensity $I$ as $I={1.37\times10^{18}} (a/\lambda[\mu\mathrm{m}])^2 [\mathrm{W cm}^{-2}]$,  ${\gamma = \sqrt{1+a^2}}$ is the Lorentz factor of the electrons, $\tilde{n} = (n-\bar{n})/\bar{n}$ is the normalized perturbed electron density,  and $n$ and $\bar{n}$ are the local and average electron densities, respectively. $c$ is the speed of light in vacuum and $\omega_p$ is plasma frequency at density $\bar{n}$.
The expansion in Eq.~(\ref{1}) is valid for $\tilde{n} \ll 1$ and $a\ll1$. We include a heuristic damping factor $\nu$ in Eq.~(\ref{2}) to describe plasma wave damping. 
Our discussion focuses on laser pulses with duration longer than or comparable with a plasma wavelength to avoid wakefield excitation. 

The nonlinear term proportional to $a^3$ accounts for the change of mass when the electrons are driven to near the relativistic velocity in the strong laser field. Its coefficient, $-\omega_p^2/2$, results from the first-order expansion of $\omega_p^2/\gamma=\omega_p^2/\sqrt{1+a^2} \approx \omega_p^2(1-a^2/2)$. The change of plasma frequency can, in turn, modulate the laser field. This  instantaneous mutual coupling between the laser field and electrostatic potential is investigated in a more general formalism in Refs.~\cite{Malkin_pre2020,Malkin_pre2020(2)}. It is found that the plasma-to-laser back-action yields a correction term $(\omega_p^2/2) c^2\partial_{zz}(\partial_t^2 + \omega_p^2)^{-1}$ in additional to the nonlinear coefficient, $-\omega_p^2/2$. The correction term is, nevertheless, very small in tenuous plasmas, and we neglect it in the rest of this article. We will present particle-in-cell (PIC) simulations in Sec.~\ref{sec:PIC} to justify the use of Eqs.~(\ref{1})-(\ref{2}). 

The plasma wave is most strongly excited at its eigen frequency $\omega_p$, which scatters the laser wave into discrete frequencies. We therefore expand the laser field and plasma density perturbation as
\begin{align}
	a &= \sum_j a_j e^{-i\omega_j t + i k_j z} + c.c., \label{3} \\
	\tilde{n} &= \delta n e^{-i\omega_p t +i k_p z} + c.c.
\end{align} 
where $\omega_j = \omega_0 + j\omega_p >0$, $k_j = k_0 + j k_p>0$, and $\omega_0 (k_0)$ is the frequency (wavenumber) of the pump laser. 
It is worth noting that plasma density perturbation could also exhibit higher harmonics in the Langmuir wave breaking limit~\cite{Dawson_PR1959,McKinstrie_PF1984} or even below this limit if the plasma has finite temperature~\cite{Balakin_PoP2018}. We refrain from these effects as we work with very low plasma density at zero temperature.

\begin{figure}[t]
	\includegraphics[width=0.95\linewidth]{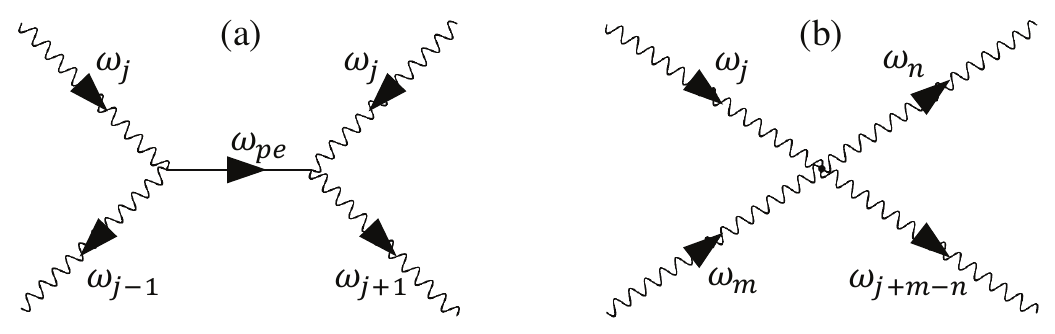}
	\caption{ Diagrams of (a) FRS and (b) FWM processes.} 
	\label{diag}
\end{figure}

In the low plasma density limit ${\omega_j\gg \omega_p}$, the dispersion relation $\omega_j^2 = c^2k_j^2 + \omega_p^2$ is approximated as $\omega_j\approx ck_j$, so $\omega_p\approx ck_p$. Matching the fast oscillation components in Eqs.~(\ref{1})-(\ref{2}) yields 
\begin{align}
	&(\partial_t + v_{g,j} \partial_z) a_j = -\frac{\omega_p^2}{2i\omega_j} (\delta n^* a_{j+1} + \delta n a_{j-1}) \nonumber \\
	&\quad + \frac{3\omega_p^2}{4i \omega_j} \left[ \sum_{k} |a_k|^2 a_j + \sum_{m,n}^{m\neq n} a_m^* a_n a_{m-n+j} + \mathrm{TSFG} \right ], \label{4} \\
	&(\partial_t + i\nu) \delta n = -\frac{\omega_p}{2i } \sum_j a_{j+1}a_j^*, \label{5}
\end{align}
where ${v_{g,j} = c^2k_j/\omega_j}$ is the group velocity of frequency component $\omega_j$. The first term on right hand side of Eq.~(\ref{4}) describes the laser anti-Stokes and Stokes Raman scattering by the plasma wave. A photon is created either by annihilation of an anti-Stokes photon and creation of a phonon or by annihilation of both a Stokes photon and a phonon, as illustrated in Fig.~\ref{diag}(a). 
Inside the square brackets are the terms for phase modulation (including self-phase modulation for $k=j$, and cross-phase modulation for $k\neq j$), FWM, and third-order sum-frequency generation (TSFG), respectively. The phase modulation nonlinearity only induces a phase shift without causing energy dissipation or redistribution. The FWM nonlinearity describes annihilation of two photons to create two new photons at different frequencies, as illustrated in Fig.~\ref{diag}(b). The FWM nonlinearity can create photons at new frequencies, similar to the FRS process, without exciting plasma waves. Each FWM stage is similar to the process discussed in Ref.~\cite{Malkin_pre2020(2)} in absence of nonlinear resonance broadening effects.
The TSFG nonlinearity describes combination of three photons and create a photon at the three-photon-sum frequency. The four photons in TSFG have drastically different frequencies making it difficult to satisfy the phase matching condition. Hence, we neglect the TSFG term in the following discussion.  Equation~(\ref{5}) describes the plasma wave amplitude driven by the beat of two adjacent electromagnetic sidebands. A phonon is created when a photon is converted into a Stokes photon.

For spatial-temporal analysis, we next introduce the coordinate system that is comoving with the speed of light: $\zeta=t-z/c$ and $\tau = z/c$. Then Eqs.~(\ref{4})-(\ref{5}) are transformed into 
\begin{align}
	&\bigg[\partial_\tau + \bigg(\frac{c}{v_{g,j}} - 1\bigg) \partial_\zeta \bigg] a_j = -\frac{\omega_p^2}{2i ck_j} (\delta n^* a_{j+1} + \delta n a_{j-1}) \nonumber \\
	&\qquad + \frac{3\omega_p^2}{4i ck_j}  \sum_l \chi_l a_{j-l} , \label{6} \\
	&(\partial_\zeta + i\nu) \delta n = -\frac{\omega_p}{2i } \sum_j a_{j+1}a_j^*, \label{7} \\
	& \chi_l = \sum_j a_{j+l} a_j^*. \label{8}
\end{align}
Here, we introduce the virtual phonon parameter $\chi_l$ to describe the phase modulation nonlinearity for $l=0$ and the FWM nonlinearity for $l\neq 0$. It obeys the relation $\chi_l^* = \chi_{-l}$. Note that the virtual phonon is not a wave, so it does not propagate or decay. 

Although the phase modulation process does not generate new discrete frequency sidebands, it flattens each sideband by inducing a chirp. For short laser pulses, the phase modulation, which is proportional to $\chi_0(\tau,\zeta)$, varies rapidly within the pulse duration. It broadens the spectrum by an amount of $[\omega_p^2\tau/(2 \omega_j)] (\partial_\zeta \chi_0$). We will show in Sec.~\ref{sec:regimes}C that $\chi_0$ remains quasi-constant during the interaction. Each sideband thus expands linearly with $\tau$, and eventually merges with adjacent sidebands when broadens to $\omega_p$. For an input pulse with duration $T$ and a peak value $\chi_{0M}$, The spectrum loses discreteness and becomes a supercontinuum  when 
\begin{equation}
	\omega_p\tau \gtrsim  \omega_jT\big/\chi_{0M}. 
\end{equation}

\begin{figure*}[th]
	\includegraphics[width=0.45\linewidth]{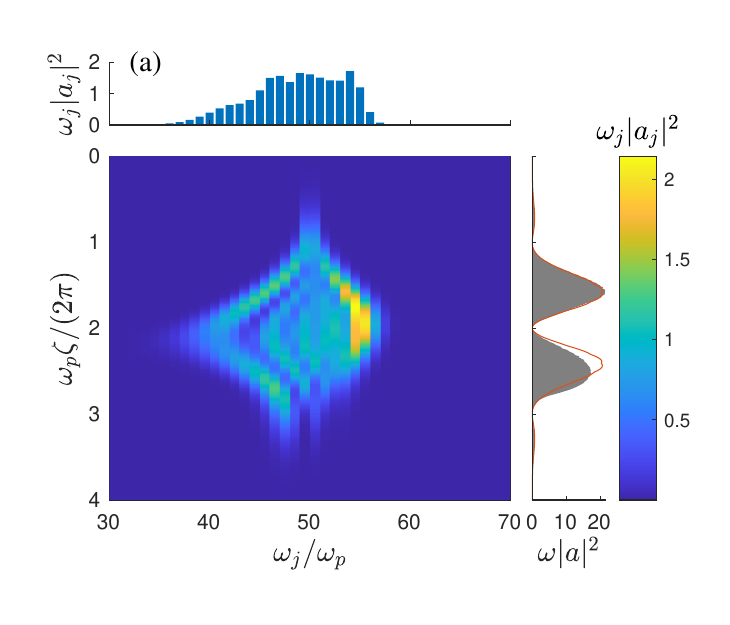}
	\hspace{3em}
	\includegraphics[width=0.45\linewidth]{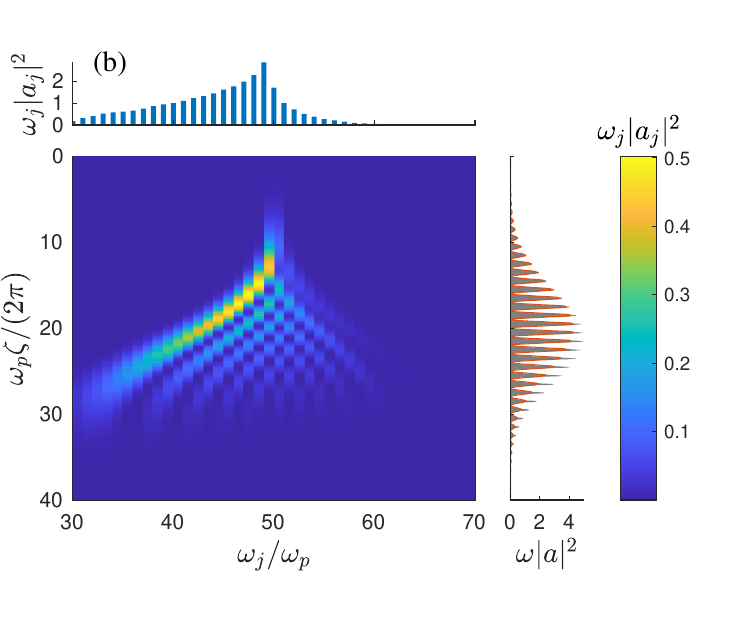}
	\caption{ (a) Tempo-spectral diagram of a short ($T=2\pi/\omega_p$) laser pulse after propagating in plasma for $250$ plasma wavelengths. The initial pulse includes components $a_{49}=a_{50}=0.4 e^{-(\zeta/T)^2}$. (b) Temporal-spectral diagram of the output of a long ($T=40\pi/\omega_p$) laser pulse after propagating in plasma for $800$ plasma wavelengths. The initial pulse includes components $a_{49}=a_{50}=0.15 e^{-(\zeta/T)^2}$. Note the difference of the vertical axis scales. The top panels show the spectra histograms for the whole pulse. The side panels show the temporal envelopes of the initial pulse (red curves) and the final pulse (gray shades). Their full-size envelope plots are shown in Fig.~\ref{FWMenv} and Fig.~\ref{FRSenv}, respectively.  } 
	\label{simu}
\end{figure*}

\section{Two regimes of operation: FWM and FRS} \label{sec:regimes}
Each new frequency component is created when an existing photon combines with a phonon or a virtual phonon. Starting with a pump photon $a_o$ and a probe photon $a_{o-1}$ at adjacent frequencies, FWM directly creates the anti-Stokes sideband $a_{o+1}$ and the Stokes sideband $a_{o-2}$ via the instantaneous virtual phonon. The pump and probe waves also excite a plasma wave $\delta n$ whose amplitude increases along $\zeta$. The plasma wave then combines with the pump and probe photon to contribute to $a_{o+1}$ and $a_{o-2}$. But the mediating plasma wave $\delta n$ introduces a $\pi/2$ phase, hence the two paths of creating the new sidebands do not coherently add to each other. 

FRS and FWM scatter the laser in distinctive manners. \textit{First}, the plasma wave amplitude grows in the direction $\zeta$, but the virtual phonon amplitude is determined solely by the instantaneous laser field. Consequently, FWM causes the most significant spectrum broadening near the intensity peak of the pulse and FRS broadening happens mostly in the pulse tail. This also leads to the \textit{second} distinction that FWM is prominent only with short pulse duration relative to plasma wave periods, and otherwise FRS dominates. 
\textit{Third}, FRS can only change the frequency of a photon by a single plasma frequency $\pm \omega_p$, but FWM can cause changes in multiple plasma frequencies $\pm j\omega_p$ even for non-integer $j$'s. \textit{Fourth}, FWM creates both a low-frequency photon and a high-frequency photon simultaneously, but FRS creates Stokes or anti-Stokes photons independently. Since the interaction rate is higher for frequency downshift, FRS overall creates more low frequency photons.  

To illustrate the interaction properties of FWM and FRS, we numerically solve Eqs.~(\ref{6})-(\ref{8}) and show the results in Fig.~\ref{simu}. We first consider a short Gaussian laser pulse with rms duration $T=2\pi/\omega_p$. The input pulse contains two frequency components at $a_{49}=a_{50}=0.4 e^{-(\zeta/T)^2}$. The double-hump pulse envelope is the result of beating. Their interaction is dominated by FWM in the $250$ plasma-wavelength-long plasma. The tempo-spectral diagram in Fig.~\ref{simu}(a) shows that the maximum spectral broadening happens near the peak of the pulse. The output spectrum shown in the top bar plot has a relatively flat distribution with sharp decent in both ends. The side panel shows that the pulse temporal envelope on the retains its structure despite some slight distortion.

For comparison, we demonstrate a FRS-dominated spectral broadening interaction by increasing the laser pulse duration to $T=40\pi/\omega_p$. The pulse amplitude is correspondingly reduced to $a_{49}=a_{50}=0.15 e^{-(\zeta/T)^2}$ and the total propagation distance is $800$ plasma wavelengths. The plasma wave decay $\nu$ is neglected. The output tempo-spectral diagram in Fig.~\ref{simu}(b) shows increasingly broader spectrum towards the tail of the laser pulse. It forms a grid-like structure with alternating dark and bright spots in both $\zeta$ and $\omega$ directions. It indicates that, for a certain frequency component, the pulse is transformed into a series of pulse trains. The time-integrated spectrum, as illustrated in the top bar plot, shows that most of the photons are shifted to lower frequencies. The spectrum has a constant decreasing trend in towards both lower and higher frequency limits. The temporal envelope retains its modulation structure despite higher peak amplitudes in the pulse tail.

\subsection{Conservation of energy and photon number}
The downshift of photon frequencies in the FRS frequency comb indicates a loss of laser energy, which seems to be conserved in the FWM-dominated frequency comb, as shown in Fig.~\ref{simu}. For more rigorous analysis, we quantitatively investigate Eqs.~(\ref{6})-(\ref{8}). First, we remind ourselves that the photon energy density in plasma can be expressed as $\mathcal{E}_\mathrm{EM} = \{\partial_\omega[\omega\epsilon(\omega)] \epsilon_0 E^2 + B^2/\mu_0\}/2 = \epsilon_0 E^2 \propto \omega^2a^2$ according to the Landau-Lifshits formula. The photon number density hence is $\mathcal{E}_\mathrm{EM}/\omega \propto \omega a^2$. Note that these two expressions have a different form in vacuum. The conservation of total photon number can be obtained exactly from Eq.~(\ref{6}), \ie $ \partial_\tau \left[\int \sum\omega_j|a_j|^2 \md\zeta \right] =0$, where $\omega_j|a_j|^2$ represents the photon number density of the component with frequency $\omega_j$ in plasma. 

For local photon number density and local laser energy density, we obtain from Eqs.~(\ref{6})-(\ref{8}) that
\begin{align}
	&\partial_\tau \sum_j \omega_j |a_j|^2 = \partial_\zeta \sum_j (1-\frac{c}{v_{g,j}})\omega_j |a_j|^2, \label{9} \\
	&\partial_\tau \sum_j \omega_j^2 |a_j|^2 = \partial_\zeta \sum_j (1-\frac{c}{v_{g,j}})\omega_j^2 |a_j|^2 - \omega_p^2  \partial_\zeta |\delta n|^2. \label{10}
\end{align} 
We see that the photon number density changes only due to group velocity dispersion (GVD), but the laser energy density is affected by both GVD and FRS. Equation~(\ref{10}) has a form of $\partial_\tau \mathcal{E}_\mathrm{EM} = (1/c)\partial_\zeta \mathcal{S}$, which describes the convection of local laser energy $\mathcal{E}_\mathrm{EM} = \partial_\tau \sum_j \omega_j |a_j|^2$ out of the local region by GVD and by conversion into plasma waves. By integrating over the laser pulse duration from $\zeta = 0$ to $\zeta_f$, we find the laser energy dissipation rate $ \partial_\tau \int_0^{\zeta_f} \mathcal{E}_\mathrm{EM} \md\zeta = \omega_p^2  |\delta n(\zeta_f)|^2$ is exactly the plasma wave energy density at the pulse tail. Although the plasma wave transports energy towards the tail of the laser pulse, FRS does not lead to spreading of photon number density, as shown in Eq.~(\ref{9}).  

Remarkably, the parametric phase modulation and FWM nonlinearities do not play a role in either photon number density redistribution or laser energy density dissipation. They only contribute to the spectrum broadening by affecting the optical refractive index of local plasmas: The plasma electrons are driven to near relativistic speed by the strong laser field and begin to oscillate in an anharmonic manner. The anharmonicity induces optical nonlinear interaction among different laser frequency components. This parametric nonlinear process does not induce any growing plasma density perturbation or electrostatic fields, and hence does not cause laser energy dissipation. But it changes plasma dispersion relation and induces phase change to local photons. In another word, FWM induces virtual phonons and FRS induces real phonons. 


\subsection{FRS and growth of plasma waves} \label{sec:phonon}
Broadening of the laser spectrum is mediated by phonons and virtual phonons of finite amplitudes. With a multi-color input laser, the virtual phonon amplitude becomes nonzero instantaneously, but the phonon amplitude grows gradually. To analyze their growth, we separate the different regimes of interaction depending on the pulse duration.

For pulses with duration longer than a plasma wavelength, FRS  dominates. We neglect the FWM interaction and find after combining Eqs.~(\ref{6})-(\ref{7}) that 
\begin{align} \label{11}
	 (\partial_\zeta - i\nu) \partial_\tau \delta n &= \partial_\zeta \sum_j (2-\frac{c}{v_{g,j}} -\frac{c}{v_{g,j+1}}) a_j^*a_{j+1} \nonumber \\
	& + \frac{\omega_p^3}{4 } \sum_j \left(\frac{1}{\omega_{j-1}} - \frac{1}{\omega_{j+1}} \right) |a_j|^2 \delta n  . 
\end{align}
The two terms on the right hand side describe the effect of GVD and the growth of the plasma wave, respectively. With plasma decay neglected, the second term indicates that the growth of plasma waves is caused by asymmetric interaction rates of phonon absorption and phonon creation: 
Each laser photon $a_j$ can emit a phonon by converting itself into a Stokes photon or absorbing a phonon by converting into an anti-Stokes photon. Since the coupling strength is larger for lower frequency components ($\omega_{j-1}^{-1} > \omega_{j+1}^{-1}$), a photon is more likely to be down-converted and create a phonon. The increased phonon number,  in turn, enhances the photon-photon interaction. 

\begin{figure}[t]
	\includegraphics[width=0.95\linewidth]{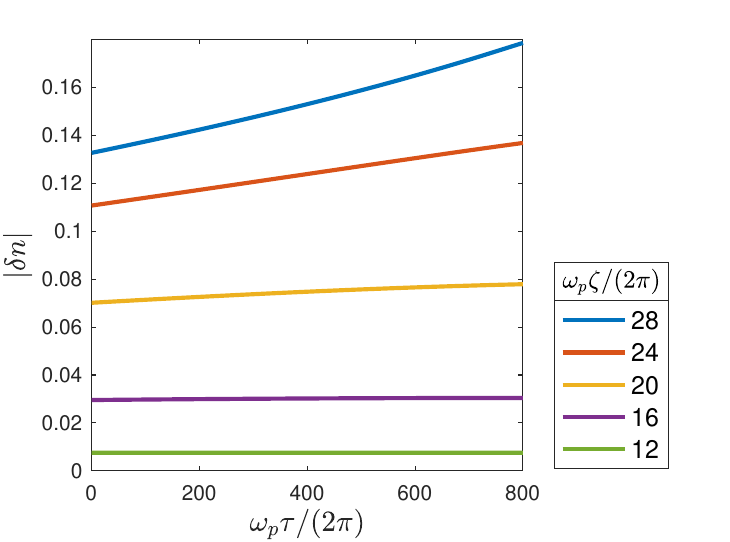} 
	\caption{ Plasma wave amplitude at various locations $\zeta$ and pulse propagation distances $\tau$. The parameters are the same with Fig.~\ref{simu}(b).} 
	\label{phonon}
\end{figure}

To find the growth rate, we neglect GVD and write $\partial_{\zeta\tau} \delta n = (\omega_p^3/4) \sum_j (\omega_{j-1}^{-1} - \omega_{j+1}^{-1}) |a_j|^2 \delta n \approx \sum_j(\omega_p^4/\omega_j^2) |a_j|^2 \delta n/2$. Since the created plasma wave does not propagate, we neglect its spatial dynamics, \ie $\partial_{tt} \delta n \approx \partial_{\zeta\tau} \delta n$. Therefore, the plasma wave growth rate in the long pulse limit is $\Gamma_\mathrm{FRS,4} \cong (\omega_p^2/\sqrt2) \left(\sum_j |a_j|^2/\omega_j^2\right)^{1/2}$, assuming $\nu\to0$. It can be reduced to $\omega_p^2 a_0/(\sqrt2 \omega_0)$ in the case of a monochromatic pump. It agrees with Refs.\cite{Mori_prl1994, Decker_1996PoP, Decker_pop19962}, apart from a factor of two difference due to the definition of $a_j$ in Eq.~(\ref{3}). The phonon peak amplitude locates at $\zeta=\tau/2$, \ie $z=ct/2$. For laser pulses with a short duration $T$, the plasma wave grows exponentially only for a finite time $T$. The plasma wave amplitude reaches its maximum amplitude at distance $\zeta$, and grows as $\exp\left[\Gamma_\mathrm{FRS,4} \sqrt{\tau\zeta}\right]$. which is shown in Fig.~\ref{phonon}.

For high plasma density ($\omega_{j-1}^{-1} \gg \omega_{j+1}^{-1}$), the laser interaction with its anti-Stokes wave becomes negligible compared to interaction with its Stokes wave. Then, the phonon dynamics can be simply described by $\partial_{tt} \delta n \approx \partial_{\zeta\tau} \delta n = [\omega_p^3/(\omega_0-\omega_p)]  |a_0|^2 \delta n/4 $. Thus, we obtain the ``three-wave'' FRS growth rate  $\Gamma_\mathrm{FRS,3} = [\omega_p^3/(\omega_0-\omega_p)]^\frac12 (a_0/2)$. Since the ``three-wave'' coupling does not consider the generation  of anti-Stokes wave which absorbs phonons, $\Gamma_\mathrm{FRS,3}$ is larger than the  ``four-wave'' FRS growth rate $\Gamma_\mathrm{FRS,4} $. 

When a phonon interacts with a photon $a_j$, whether the scattering creates an anti-Stokes photon $a_{j+1}$ or a Stokes photon $a_{j-1}$ depends on the relative phase of $\delta n$ and $a_j$. We obtain the photon number dynamics from Eq.~(\ref{6}) that
\begin{align}
	&\quad [\partial_\tau + (\frac{c}{v_{g,j}}-1) \partial_\zeta] (\omega_j |a_j|^2) \nonumber \\
	&=  \mathrm{Re} \big[ \omega_p^3 (a_{j+1}^*\delta n -  a_{j-1}^*\delta n^*)a_j  \big]  - \mathrm{Im} \big[ 3\omega_p^2  \sum_{k} a_ja_{j+k}^*\chi_k\big], \label{12}
\end{align}
and $\delta n = \int_0^\zeta \chi_1 e^{-\nu\zeta'}\md\zeta' $.
On the right hand side, the first term describes the FRS interaction and the second term describes the FWM interaction. 
The FRS interaction can be categorized into resonant terms and non-resonant terms. 

The resonant photon number growth due to FRS interaction is proportional to $(|a_{j+1}|^2-|a_{j-1}|^2) |a_j|^2$ with the $\zeta$ dependence neglected. In the small time scale, it leads to an exponential growth of the photon number in mode $a_j$. This amplitude-difference-driven interaction causes a cascade of photon frequency decrease: 
Starting with a bi-color laser input $a_o$ and $a_{o-1}$, the resonant FRS interaction initially causes $|a_o|$ to decrease and $|a_{o-1}|$ to increase. A low-frequency mode $|a_{o-2}|$ is created and grows. As $|a_{o-2}|$ approaches and exceeds $|a_o|$, the mode $|a_{o-1}|$ begins to decrease. Without photon supplementation from higher frequency modes, $|a_o|$ eventually reaches zero amplitude. Overall, modes are created at the low-frequency limit and are annihilated in the high-frequency limit, causing a successive frequency downshift. The trend of downshift can be seen as the bright stream in main plot of Fig.~\ref{simu}(b). The resonant frequency downshift process asymptotically results in more low-frequency sidebands with monotonically decreasing amplitudes, as can be seen in the top panel of Fig.~\ref{simu}(b).

The non-resonant FRS interaction is described by the terms proportional to $Re[(a_{j+1}^* \int_0^\zeta\chi_1 e^{-\nu\zeta'}\md\zeta' - a_{j-1}^* \int_0^\zeta\chi_1^* e^{-\nu\zeta'}\md\zeta') a_j]$. Since it is proportional to $a_j$, the non-resonant FRS has a lower growth rate. The non-resonant FRS can create both frequency upshifted photons and frequency downshifted photons, similar to FWM-type interaction.

\begin{figure}[b]
	\includegraphics[width=0.95\linewidth]{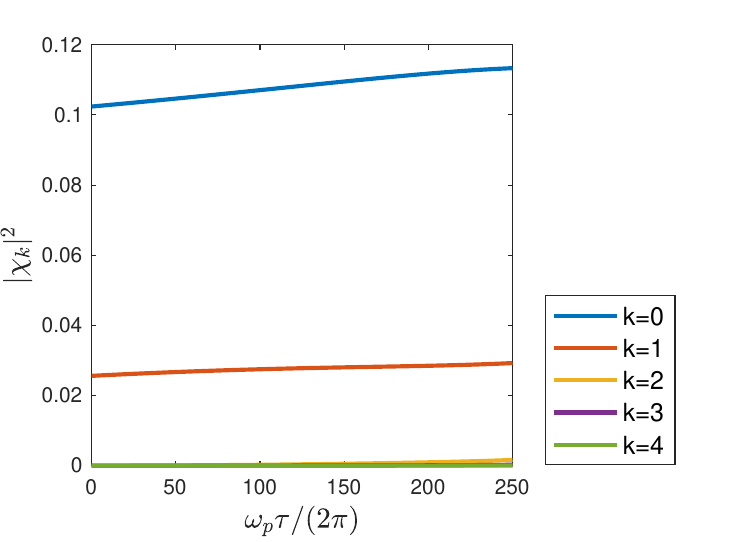} 
	\caption{ Amplitudes of virtual phonon $|\chi_k|$ at the pulse peak $\zeta/(c\omega_p)=1$. The parameters are the same with Fig.~\ref{simu}(a).} 
	\label{chi}
\end{figure}

\subsection{FWM and evolution of virtual phonons}
Laser pulses with a short duration near a plasma wavelength are not sufficient to generate strong plasma waves and hence FWM dominates. FWM creates new frequency components by scattering the laser photons via virtual phonons. The virtual phonons do not have an eigenfrequency and are totally determined by the beating lasers. Hence, FWM can upshift or downshift a laser photon by multiple times the plasma frequency.

The FWM growth rate depends on the virtual phonon amplitude. 
With neglected FRS interaction, the virtual phonon dynamics is obtained by combining Eqs.~(\ref{6}) and (\ref{8}) that 
\begin{align}
	& \partial_\tau \sum_k |\chi_k|^2 = \sum_{j,k}\chi_k (1-\frac{c}{v_{g,j}})\partial_\zeta (a_ja_{j+k}^* + a_{j-k}a_j^*), \label{13} \\
	& \partial_\tau (\chi_k \pm \chi_{-k}) = \sum_j (1-\frac{c}{v_{g,j}})\partial_\zeta [a_j(a_{j+k}^* \pm a_{j-k}^*)] \nonumber \\
	&\qquad + \frac{3\omega_p^2}{4i} \sum_{j,k'} \left(\frac{1}{\omega_j} - \frac{1}{\omega_{j-k-k'}} \right)\chi_{k'}a_{j-k'}a_{j-k}^* \nonumber \\
	&\qquad \pm \frac{3\omega_p^2}{4i} \sum_{j,k'} \left(\frac{1}{\omega_j} - \frac{1}{\omega_{j+k-k'}} \right)\chi_{k'}a_{j-k'}a_{j+k}^*. \label{14} 
\end{align}
Since $\chi_{-k}=\chi_k^*$, Eq.~(\ref{14}) for $\pm$ describes the evolution of the real and imaginary parts of $\chi_k$, respectively. 
The identity (\ref{13}) indicates that the total virtual phonon amplitude is conserved when neglecting GVD. Hence, none of $\chi_k$ can grow exponentially. With GVD, growth of $\chi_k$ is driven by other $\chi_{k'}$ ($k'\neq k$) terms provided that $a_ja^*_{j-(k-k')}$ is nonzeros. Since the coefficients $(1/\omega_j-1/\omega_{j\pm k-k'}) \sim (k'\pm k)/(j\omega_j)$ are small values, the growth of $\chi_k$ is lower than the growth of $a_j$. Hence, $\chi_k$ can be approximated as quasi-constants. The quasi-conservation is numerically verified and illustrated in Fig.~\ref{chi} which shows only nonzero $\chi_{0,\pm1}$ for the bi-color input laser.

The photon number growth due to FWM is described by the second term of Eq.~(\ref{12}). Due to the $\mathrm{Im}$ operation, it does not include any resonant terms. Thus, the spectrum expands to both lower and higher frequencies equally. Different from FRS interaction, FWM grows fastest at the pulse intensity peak. Thus, the tempo-spectral diagram in Fig.~\ref{simu}(a) shows broad bands only at the peak center. 

The dominance of low-order terms $\chi_{0,\pm1}$ in Fig.~\ref{chi} means that the new laser frequency components are generated at an interval of $\omega_p$. This differs from Ref.~\cite{Malkin_pre2020(2)} which aims at substantially upshifting the laser frequency by injecting two highly detuned pulses with frequency differences greater than the plasma frequency. Similar to Ref.~\cite{Malkin_pre2020(2)}, however, both upshift and downshift of the photon frequency coexist in the FWM process we describe here.



\section{Scaling of frequency bandwidth growth} \label{sec:scaling}
With a bi-color laser input, only $\chi_0$ and $\chi_{\pm1}$ are nonzero. 
For quasi-steady values of $\chi_k$, we can find the analytical solution to the recursion equation in the limit of small bandwidth $\mathrm{max}(\omega_j)-\mathrm{min}(\omega_j) \ll \omega_j$, and hence 
\begin{multline}
	a_j(\tau,\zeta) = \sum_o a_o(0,\zeta)e^{i(j-o)(\frac{\pi}{2}-\psi)} \\
 \times \exp\left(\frac{3i\omega_p^2\chi_0\tau}{4\omega_j} \right) J_{j-o}\bigg(\frac{3\omega_p^2|\chi_1|}{2\omega_o }\tau \bigg), \label{15}
\end{multline}
where $o$'s denote the indices of the input laser fields, and $\psi$ is the phase of $\chi_1$, \ie $e^{i\psi} = \chi_1/|\chi_1| \cong 1$. The solution shows that each input laser component expands to a broad spectrum whose amplitude is described by the Bessel function $J_{j-o}$. The contribution from different $a_o$'s differ by a phase of $\pi/2$, so they do not interfere. 
The photon number in each mode is then
\begin{equation}
	\omega_j |a_j(\tau,\zeta)|^2 \cong \omega_j\sum_o |a_o(0,\zeta)|^2 J_{j-o}^2\bigg(\frac{3\omega_p^2|\chi_1|}{2\omega_o }\tau \bigg).
\end{equation}
Since the first peak of $J_{\pm j}(x)$ locates at approximately $x\sim |j|$, laser spectral width $\Delta\omega_\mathrm{FWM}$ expands with the scaling of 
\begin{equation} \label{17}
	\Delta\omega_\mathrm{FWM} \sim \frac{3\omega_p^2|\chi_1|}{\omega_o }\tau = \frac{3\omega_p^2|a_oa_{o-1}^*|}{\omega_o }\tau. 
\end{equation}
For the parameters used in Fig.~\ref{simu}(a), the output bandwidth reaches $\pm 16\omega_p$, which agrees well with the simulation results.

Equation~(\ref{15}) has a similar form with the solution of FRS interaction found by Karttunen and Salomaa~\cite{Salomaa_1986,KarttunenPRL1986} augmented with the self-phase modulation term. For FRS interaction, Eq.~(\ref{15}) is to be modified by replacing $e^{i\psi} = \delta n/|\delta n| \cong i$ and $|\chi_1|/  \to |\delta n|$ 
\begin{equation}
	a_j(\tau,\zeta) \simeq \sum_o a_o(0,\zeta)(-1)^{j-o}  J_{j-o}\bigg(\frac{\omega_p^2|\delta n|}{\omega_o}\tau \bigg), \label{18}
\end{equation}
The change of $\psi$ causes a $\pi$ phase difference between the spectra from different $a_o$'s. Using the identity $J_{-j}(\cdot) = (-1)^j J_j(\cdot)$, we find the estimation of the photon number
\begin{equation}
	\omega_j |a_j(\tau,\zeta)|^2 \simeq \omega_j \left[ \sum_o (\mp1)^o |a_o(0,\zeta)| J_{j-o}\bigg(\frac{\omega_p^2|\delta n|}{\omega_o}\tau \bigg) \right]^2, 
\end{equation}
where $\mp$ are for positive and negative values of $j-o$, respectively. Thus, the spectral broadening to the high frequency band is suppressed due to destructive interfere; The spectral expansion to the lower frequency band is enhanced due to constructive interference, as seen from Fig.~\ref{simu}(b). Since the spectral broadens only to the lower frequency bands, the scaling is
\begin{equation} \label{20}
	\Delta\omega_\mathrm{FRS} \sim \frac{\omega_p^2|\delta n|}{\omega_o}\tau = \frac{\omega_p^3\tau}{2\omega_o } \left| \int_0^\zeta a_oa_{o-1}^* e^{-\nu\zeta'}\md\zeta'  \right|.
\end{equation}
For the parameters used in Fig.~\ref{simu}(a), the output spectrum extends to a lower frequency by $\pm 15\omega_p$, which agrees in the order of magnitude with the simulation results. 
It should be born in mind that the solution only approximately describes the spectral evolution because the plasma amplitude $\delta n$ changes as we see from Fig.~\ref{phonon}. 

Comparing Eqs.~(\ref{17}) and (\ref{20}), we find that the spectral bandwidth broadening due to FWM or FRS have the same dependence on the parameters including interaction time $\tau$ and input frequency $\omega_o$. They both increase with higher pump amplitude multiplication $\chi_1(0,\zeta) = a_o(0,\zeta)a_{o-1}(0,\zeta)$, but FWM grows proportionally to $|\chi_1(0,\zeta)|$ and FRS depends on its integration $\big| \int_0^\zeta \chi_1(0,\zeta') e^{-\nu\zeta'}\md\zeta' \big|$. Therefore, FWM dominates when the pulse duration is as short as a few plasma wavelengths and it broadens the laser spectrum most significantly at the pulse intensity peak; FRS dominates with longer pulse duration and it broadens the laser spectrum most significantly at the pulse tail.

\section{Pulse envelope modulation} \label{sec:envelope}
A larger laser spectral bandwidth $\Delta\omega$, in principle, supports shorter laser pulses provided that all the frequency components constructively interference with the identical phase. However, Eqs.~(\ref{15}) and (\ref{18}) show that the frequency comb components have both opposite phases at different $j$'s. The phase flipping prevents the pulse from forming sharp peaks.

\begin{figure}[t]
	\includegraphics[width=0.95\linewidth]{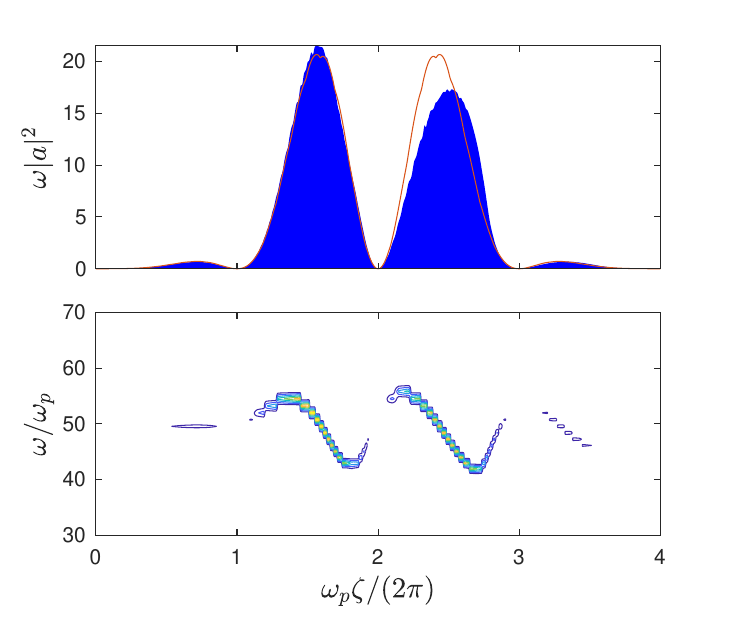} 
	\caption{ Top panel: The input (red curve) and output (blue shade) pulse envelopes of a short laser pulse $T=2\pi/\omega_p$. Bottom panel: contour plot of the time-frequency synchrosqueezed transform. The parameters are the same with Fig.~\ref{simu}(a).} 
	\label{FWMenv}
\end{figure}

\begin{figure}[t]
	\includegraphics[width=0.95\linewidth]{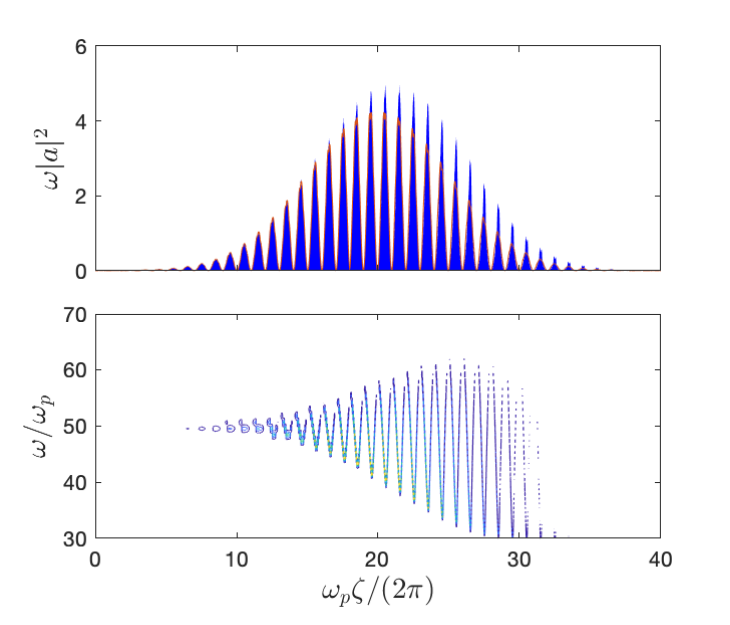} 
	\caption{ Top panel: The input (red curve) and output (blue shade) pulse envelopes of a short laser pulse $T=40\pi/\omega_p$. Bottom panel: contour plot of the time-frequency synchrosqueezed transform. The parameters are the same with Fig.~\ref{simu}(b).} 
	\label{FRSenv}
\end{figure}

Actually, we can find the analytical solution to the pulse temporal envelope in the limit of large spectral width. Note the generating function of the Bessel function, $ e^{ix\cos(\phi)} = \sum_{j=-\infty}^{+\infty}i^j J_j(x) e^{ij\phi} $. For FWM interaction, the laser field $a$ can be found by combining Eqs.~(\ref{3}) and (\ref{15}) and using $\omega_j-\omega_o=(j-o)\omega_p$ 
\begin{align} \label{21}
	a &= 2\mathrm{Re}\sum_{o,j} a_o(0,\zeta)i^{j-o} e^{\frac{3i\omega_p^2\chi_0\tau}{4\omega_j }} J_{j-o}\bigg(\frac{3\omega_p^2|\chi_1|}{2\omega_o }\tau \bigg) e^{-i\omega_j\zeta} \nonumber \\
	&\approx 2\mathrm{Re}\sum_{o} a_o(0,\zeta) e^{\frac{3i\omega_p^2\chi_0\tau}{4\omega_o } +i(\frac{3\omega_p^2|\chi_1|}{2\omega_o }\tau ) \cos(\omega_p\zeta) + i\omega_o\zeta} \nonumber \\
	&= \sum_{o} a_o(0,\zeta) \cos\left[ \frac{3\omega_p^2\chi_0}{4 \omega_j}\tau +  (\frac{3\omega_p^2|\chi_1|}{2\omega_o }\tau ) \cos(\omega_p\zeta) + \omega_o\zeta \right].
\end{align}
The result shows that each pump pulse $a_o$ is frequency modulated by the FWM interaction: The $\chi_0$ term describes the phase modulation which induces a chirp proportional to $\partial_\zeta \chi_0(\tau,\zeta)\tau  = \partial_\zeta \left| \sum_o a_o(0,\zeta) \right|^2\tau$; The $\chi_1$ term creates new sidebands and modulates the envelope at the frequency $\omega_p$. The frequency modulation index $\Delta\omega_\mathrm{FWM}/2 =\omega_p|\chi_1|\tau/(\omega_o )$ grows linearly with $\tau$. The increasing modulation index due to FWM broadens the laser spectrum thereby allowing for pulse duration compression. 


For FWM-dominated spectral broadening, the pump pulse duration is shorter than a few plasma wavelengths. The frequency modulation thus causes a frequency chirp near the pulse center $\zeta_M$. Due to the strong $\zeta$ dependence of $\chi_0(\tau,\zeta) \simeq \sum_o |a_o(0,\zeta)|^2$, the phase modulation enhances the frequency chirp by causing increasingly higher frequency upshift towards the tail. On the other hand, since higher frequency components propagates faster in plasam, GVD causes negative chirp. If the chirping by FWM and GVD is balanced, the laser pusle duration is then compressed  and the peak amplitude is enhanced. Such a principle is adopted in Refs.~\cite{Kalmykov_2005PRL,Kalmykov_2006PRE} to obtain few-cycle laser spikes. 

Figure~\ref{FWMenv} zooms in the side panel of Fig.~\ref{simu}(a) and shows the frequency chirp. The filled body represents the fast oscillating pulse envelope at $2\pi/(50\omega_p)$. Note that the double-hump structure of the input pulse arises from beating of the two frequency components. The pulse envelope maintains the same structure. 
The output pulse obviously develops negative frequency chirp indicating a dominating role of GVD. The nearly linearly chirped peaks could then be compressed into two sharp and intense peaks through proper dispersion management.

For FRS interaction, the laser field can be found similarly by combining Eqs.~(\ref{3}) and (\ref{18})
\begin{align} \label{21}
	a &\simeq 2\mathrm{Re}\sum_{o,j} a_o(0,\zeta)(-1)^{j-o} J_{j-o}\bigg(\frac{\omega_p^2|\delta n|}{\omega_o}\tau \bigg) e^{-i\omega_j\zeta} \nonumber \\
	&\approx 2\mathrm{Re}\sum_{o} a_o(0,\zeta) \exp\left[i\bigg(\frac{\omega_p^2|\delta n|}{\omega_o}\tau \bigg) \sin(\omega_p\zeta)  + \omega_o\zeta \right]\nonumber \\
	&= \sum_{o} a_o(0,\zeta) \cos\left[ \bigg(\frac{\omega_p^2|\delta n|}{\omega_o}\tau \bigg) \sin(\omega_p\zeta)  + \omega_o\zeta \right].
\end{align}
The result shows that FRS causes frequency modulation of the laser pulse with increasing modulation index, which is similarly to FWM. The broadened spectrum, combined with GVD, compresses the pulse duration and increases the peak amplitude.


Figure~\ref{FRSenv} zooms in the side panel of Fig.~\ref{simu}(b) and show the instantaneous frequency within the pulse. The isolated peak structure with a period of $2\pi/\omega_p$ is the result of beat between two pumps at $49\omega_p$ and $50\omega_p$. The plot exhibits the most change of pulse envelope in the peak and tail of the pulse where plasma wave is the strongest. Because chirp is developed individually within each spike, the pulse train cannot be compressed into a single pulse to increase its peak intensity.

\begin{figure*}[thb]
	\includegraphics[width=0.95\linewidth]{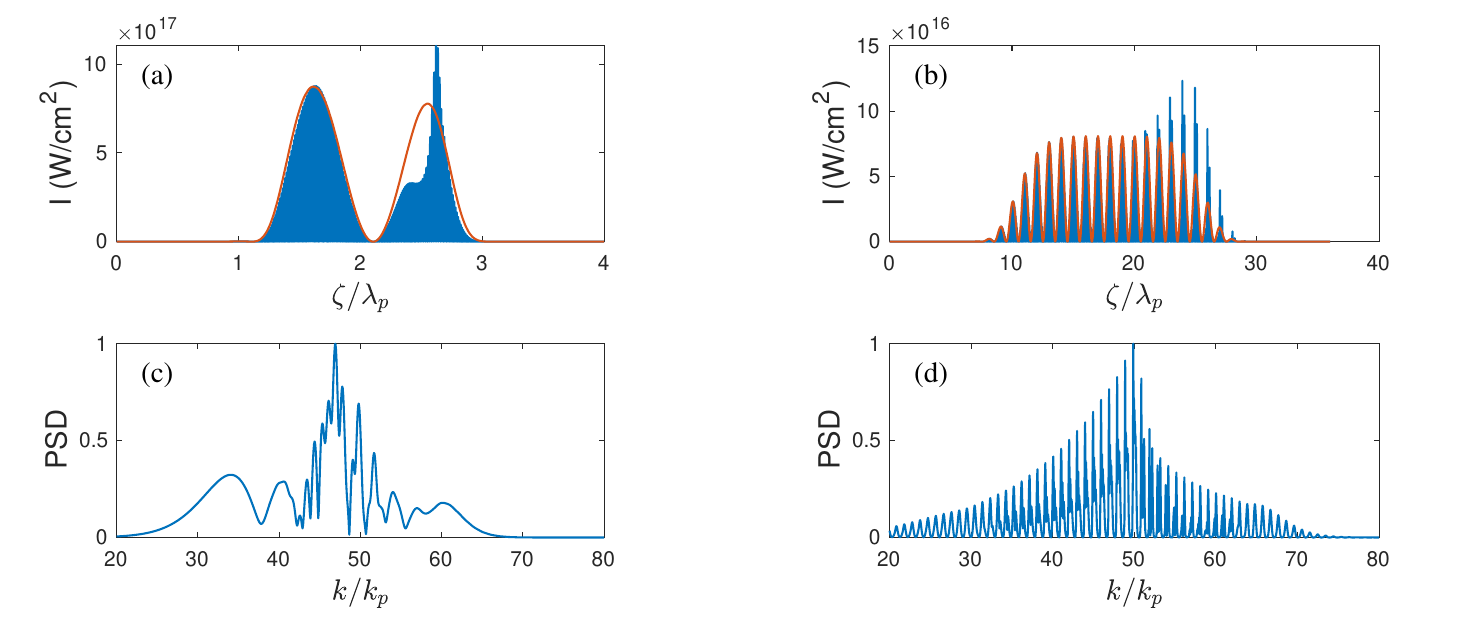} 
	\caption{PIC simulation results of a supercontinuum [(a) and (c)] generated from a short laser pulse in the FWM-dominated regime, and a frequency comb [(b) and (d)] generated from a long laser pulse in the FRS-dominated regime. In (a) and (b), the red curves show the initial pulse envelope, and the blue shades show the output signal. (c) and (d) show the power spectral density (PSD) of the corresponding output signals.  } 
	\label{PIC}
\end{figure*}

\section{PIC simulations} \label{sec:PIC}
As proof-of-principle demonstration of generating supercontinuum and frequency comb, we conduct particle-in-cell (PIC) simulations using the full-relativistic kinetic code EPOCH~\cite{EPOCH2015}.  The input laser pulse of each simulation comprises of two frequency components with wavelengths $1\,\mu$m and $0.98\,\mu$m, respectively. They both have a Gaussian profile, \ie $a=\sum_o a_o \exp[-\frac12 (\frac{\zeta}{T})^2 + i\omega_ot]$, and their beat causes multiple-peak structure. 
The uniform plasma has a density of $4.47\times10^{17}\, \mathrm{cm}^{-3}$, corresponding to a plasma frequency of $1/50$ of the $1\,\mu$m laser. Hence, the plasma wavelength is $\lambda_p=2\pi/k_p=50\,\mu$m. The output laser pulses after propagating through $40$-mm-long plasma ($800\lambda_p$) are shown in Fig.~\ref{PIC}.

Figure~\ref{PIC}(a) and (c) show frequency broadening of a short and intense pulse into a supercontinuum. Each input pulse component has duration $T=0.15\,\mathrm{ps}$ ($0.9$ plasma period) and peak amplitude $a_o=0.4$ (corresponding to $I=2.2\times10^{17}\,\mathrm{Wcm}^{-2}$). The output pulse envelope shows some degree of compression in the tail. We take Fourier transform of the laser electric field at the snapshot to obtain its wavevector spectrum. The spectrum in Fig.~\ref{PIC}(c) shows a supercontinuum with bandwidth of $\sim0.8\omega_o$.

Figure~\ref{PIC}(b) and (d) show frequency broadening of a long and less intense pulse into a frequency comb. Each input pulse component has duration $T=1.5\,$ps ($9$ plasma periods) and peak amplitude $a_o=0.12$ (corresponding to $I=2\times10^{16}\,\mathrm{Wcm}^{-2}$). The output envelope shows a small amount of pulse compression in the tail of the pulse. Fourier transform of its electric field yields a frequency comb with discrete equidistant spikes spanning from below $20k_p$ to $70k_p$. The result agrees well with our analysis, which justifies the use of Eqs.~(\ref{1})-(\ref{2}).

\section{Conclusion and Discussion} \label{sec:concl}
In conclusion, we show that a laser pulse can be expanded into a broadband spectrum when propagating through tenuous plasmas. The spectrum broadening arises from a cascade of  both Stokes and anti-Stokes scattering due to plasma waves and electron relativistic effects. We point out that a few-cycle pulse with near-relativistic intensity can produce an octave-wide supercontinuum through FWM and phase modulation; and a multi-cycle pulse can produce an octave-wide frequency comb through FRS. As the frequency comb bandwidth increases, it continues to lose energy to the plasma wave. As a result, the lower frequency components grow faster than the higher ones, and the comb loses total laser energy. The comb to expand to lower frequencies. But the supercontinuum conserves laser energy because the amplitudes of virtual phonons do not change. Hence, the supercontinuum spectrum broadens symmetrically to both lower and higher frequencies. A notable mention is that the similar FWM process is investigated in Ref.~\cite{Malkin_pre2020(2)} to achieve resonant laser frequency doubling by arranging the frequencies and intensities of two highly detuned laser. Our current article focuses on a different regime that both FRS and FWM processes broaden the input laser spectrum by multiple integer times the plasma frequency. 

Compared to using optical crystals for generating optical supercontinuum and frequency combs, plasmas have high thermal damage tolerance and can work in the near-relativistic regime. The ultra-intense broadband pulses are particularly useful for minimizing laser scattering and absorption laser-plasma application like inertial confinement fusion and laser-plasma accelerators. Using plasmas, the supercontinuums or frequency combs can be generated in the EUV or x-ray regimes. The equidistant peaks of the frequency comb could also enable potential applications in ultrafast optics at ultra-high intensities, for example, creating high-intensity THz waves. 

The experimental feasibility of this method is demonstrated through PIC simulations of the frequency comb using accessible parameters. The parameters of laser wavelength, peak intensity and plasma length are similar to those in laser particle accelerators~\cite{LPA_rmp2009}, but the requirements are less stringent for generations of supercontinuum or frequency comb. The peak laser intensity of $\unit[10^{16}-10^{17}]{W cm^{-2}}$ is sufficient to enter the mildly relativistic regime. It can pass through a few-$\unit[]{cm}$-long plasma with $\unit[10^{17}-10^{18}]{cm^{-3}}$ density. At such low density, collisional plasma damping could be neglected and the consequent low plasma wavenumber (which is proportional to $\omega_p/\omega_0$) also avoids Landau damping. 

The proposed method of generating frequency combs does not rely on an optical resonator as is required for conventional methods using, \ie a mode-locked laser. Since the laser spectrum is broadened after a single pass through the plasma, the comb quality is limited by the plasma inhomogeneity. Specifically, short-range plasma density inhomogeneities destroy the FRS resonance, reducing the efficiency of frequency band broadening. More seriously, the long-range plasma density inhomogeneity could gradually shift the FRS resonance, resulting in fluctuation of the comb repetition rate. Plasma inhomogeneity, however, does not affect the generation of the supercontinuum, which does not depend on resonance with the plasma frequency. 

Our analysis applies to 1D propagation of lasers with below-relativistic intensity and pulse duration not too much shorter than a plasma wavelength. For ultra-relativistic laser intensity, the laser-plasma interaction becomes fully nonlinear~\cite{nonlienar_prl1990, nonlienar_pra1990} and our theory is no longer valid. A pulse duration much shorter than a plasma wavelength produces a strong wakefield~\cite{LPA_rmp2009}, which can significantly alter the laser envelope evolution. To avoid pulse distortion in long plasmas, the laser power needs to be below the critical power for transverse filamentation~\cite{Antonsen_prl1992, Antonsen_pfb1993}.

\begin{acknowledgments}
	This work was supported by NNSA Grant No. DE-NA0002948.
\end{acknowledgments}  

\bibliography{freqcomb}

\end{document}